\documentclass[preprint,12pt,authoryear]{elsarticle} 
\usepackage{amssymb}

\usepackage{amsmath}
\usepackage{csquotes}
\usepackage{hyperref}

\newtheorem{definition}{\textit{Modelling Assumption}}

\journal{Studies in History and Philosophy of Modern Physics}

\begin{document}

\begin{frontmatter}

\title{Black Holes are about \textit{Quantum} Information}

\author[1]{Aude Corbeel}
\author[1,2]{Sebastian De Haro}

\affiliation[1]{organization={Institute for Theoretical Physics, University of Amsterdam},
            city={Amsterdam},
            postcode={1090 GL}, 
            country={The Netherlands}}
\affiliation[2]{organization={Institute for Logic, Language and Computation and Qusoft Research Center for Quantum Software, University
of Amsterdam},
            city={Amsterdam},
            postcode={1090 GL}, 
            country={The Netherlands}}

\affiliation{a.n.i.corbeel@uva.nl --- s.deharo@uva.nl}

\begin{abstract}
We review and analyse quantum information-theoretic approaches to black hole evaporation that purport to bring us closer to a resolution of the black hole information paradox. Our focus is on the Page-time paradox and on the contribution of the Hayden–Preskill model to understanding information release under the assumption of unitary evaporation. In this model, the black hole is taken to be a finite-dimensional quantum system whose dynamics can be represented by a quantum circuit, and its information-processing and communication abilities are studied. We identify the assumptions underlying this modelling and highlight two key features, fast scrambling dynamics and entanglement as a resource, that underpin the information release in the model. We argue that this reasoning exemplifies a broader change in how information is used in black hole physics, in which quantum information theory now provides a useful operational modelling framework. This application avoids earlier philosophical objections insofar as it is not based on epistemic ignorance and does not regard information as a substance. This line of reasoning has become central to the debate and, as such, warrants greater attention from the philosophy of physics community.

\end{abstract}

\end{frontmatter}

\newpage

\tableofcontents

\section{Introduction}

Recent years have witnessed a wave of research on various aspects of black holes, giving rise to what may be called an emerging `philosophy of black holes'. Among the questions addressed, the black hole information paradox, along with related issues such as black hole thermodynamics, are of particular significance. Some work has examined the sense in which the black hole information paradox is or is not paradoxical (see e.g.~\citet{maudlin2017information,wallace2017why}), while other work has focussed on highlighting explicit and implicit assumptions in the latest incarnation of this paradox \citep{cinti2025peeking}. Additional studies have explored the thermal properties of black holes, investigating whether the relation between black hole thermodynamics and thermodynamics is more than an analogy: see for example \citet{curiel2014classical,dougherty2017blackhole,wallace2018case,wallace2019case}. This line of research has also explored whether the Bekenstein-Hawking entropy of a black hole can be interpreted as an information-theoretic quantity, especially in the framework of Shannon's information theory, as \citet{bekenstein1973entropy} initially proposed (see \citet{dougherty2017blackhole,wuthrich2017are,prunkl2019blackhole}). The criticism here is that information is an epistemic concept, while thermodynamical entropy pertains to the physical process of heat transfer at a given temperature. 

To the best of our knowledge, the philosophy of physics literature has so far not engaged with recent and more detailed \textit{quantum} information-theoretic studies of black holes in physics, particularly those that apply detailed ideas from quantum computation (such as quantum error correction) and quantum communication (such as entanglement-assisted communication) to black holes.\footnote{The interpretation of the AdS/CFT correspondence as a quantum error-correcting code, proposed by \citet{almheiri2015bulk, pastawski2015holographic}, has been addressed by \citet{bain2020spacetime}.} Instead, prior philosophical work by \citet{dougherty2017blackhole} and \citet{wuthrich2017are} has been limited to critiques of the application of \textit{classical} information theory to classical black holes.

In this paper, we aim to fill this gap by first introducing, and then analysing, the ground-breaking work of \citet{hayden2007blackholes}. There are three related reasons why we consider it worthwhile to undertake this task.

First, quantum information theory, which addresses quantitative questions about the acquisition, transmission, and processing of information in quantum systems, offers useful conceptual and technical tools for studying black holes. We will argue that these tools can help refine our understanding of the version of the information paradox at issue in this paper, namely, the Page time paradox \citep{page1993information}. The discussion surrounding this paradox extends beyond the question of unitarity, and this is precisely where the analysis by Hayden and Preskill offers new insight.

Second, Hayden and Preskill's engagement with quantum information theory allows them to study a dynamical black hole as an object that effectively performs a computation, processing information without destroying it. We argue that this approach aligns with the view articulated by Christopher Timpson in his book `Quantum Information Theory and the Foundations of Quantum Mechanics' \citep{timpson2013quantum}. His idea is that quantum information theory can serve as an alternative framework of enquiry, supplementing our fundamental physical theories, for exploring certain interesting features of material objects; notably, their capacity to store, process, and transmit information \cite[p.~237]{timpson2013quantum}. Our claim is that the application of this information-theoretic lens to black holes does not rely on epistemic reasoning and helps to counter certain forms of scepticism about the use of the concept of `information' in black hole physics.

Third, while Hayden and Preskill were not the first to apply quantum information-theoretic ideas to black hole physics, their work significantly advanced this perspective and has been taken as a major clue for later investigations. Papers such as \citet{almheiri2019entropy, penington2020entanglement}, widely seen as crucial to resolving the Page time paradox, built on the Hayden–Preskill framework by implementing it in holographic settings. Their work also played a pivotal role in exposing another aspect of this paradox: the firewall paradox \citep{almheiri2013black}, which notably arises from the possibility that an infalling observer could extract information from Hawking radiation before crossing the horizon. It has further motivated experimental investigations on quantum computers (e.g.~\citet{landsman2019verified,jafferis2022traversable}). A careful analysis of their original work is therefore essential for assessing both the strengths and limitations of the research it informed and inspired.

In more detail, \citet{hayden2007blackholes} view a black hole as playing a role in a communication system.  They consider a small amount of quantum information (namely, the information in Alice's diary), thrown into a black hole, and model its evolution using a quantum circuit, a sequence of quantum gates that abstracts away from the complex physical interactions involved in the corresponding physical system. Because of its relative simplicity and precision, this kind of model is well suited to address conceptual questions. Notably, this perspective prompts two well-defined questions about the release of information into the radiation:
\begin{itemize}
    \item[(i)] Physical mechanism of information release: \textit{How does a small amount of quantum information escape from the black hole?}
    \item[(ii)] Amount of time: \textit{How long will Bob, who is outside the black hole, need to wait to be able to recover Alice's diary?}
\end{itemize}

Hayden and Preskill's model being a {\it model} means that it contains a number of simplifying assumptions and idealisations that are good approximations to some of the behaviours of the black hole. Thus, a major part of our analysis will be to state nine modelling assumptions that go into answering questions (i) and (ii) above. In this way, we aim to clarify what is required for the model to give a successful description of a black hole. In particular, we will identify two distinct roles that entanglement plays in the putative mechanism through which information can be released: first, as a means of protecting
information against local errors via scrambling; and second, as a consumable resource that facilitates
the transfer of information. 

The central assumption on which this model relies is known as the `central dogma' of black hole physics \citep{almheiri2021entropy}: in short, seen from the outside, a black hole is a quantum system that evolves unitarily and is made up of a large but finite 
number of degrees of freedom. Establishing this central dogma requires providing the proof of unitarity, which is not the claim made by \citet{hayden2007blackholes}. Nevertheless, the assumption renders an otherwise daunting problem more tractable and is therefore valuable for studying its conceptual implications for the release of information.

The paper is structured as follows: Section \ref{sec2} provides the conceptual and theoretical foundation for applying quantum information theory to black holes by reviewing the information controversy and discussing the central dogma. Section \ref{Page’s Model and Scrambling} begins by introducing the basic features of a quantum circuit and then presents Page’s analysis in this language. Section \ref{Hayden and Preskill's Model} analyses Hayden and Preskill's treatment of a black hole as part of a communication protocol, identifying the underlying modelling assumptions and explaining the new insights that their use of quantum information theory offers regarding black holes. Section~\ref{Conclusion} concludes.

\section{Black Hole Entropy and the Role of Information}\label{sec2}

In this section, before delving into the specifics of the approach taken by \citet{hayden2007blackholes}, we provide the necessary groundwork to justify the application of quantum information theory to the study of black holes. 

Section \ref{The Information Controversy in Black Hole Physics: an Overview} reviews the issues surrounding the use of the concept of `information’ in relation to black holes. Section \ref{The Hayden--Preskill Model and the Way Forward} outlines Hayden and Preskill’s quantum information-theoretic approach, with details explored in the following sections.
 Section \ref{The Central Dogma of Black Hole Physics} motivates the central dogma, the key assumption enabling information to play a different role. Section \ref{Potential Rebuttal from Information Loss} discusses the `Page time Paradox' as a challenge to this assumption and how recent developments may address it.

\subsection{The Information Controversy in Black Hole Physics}
\label{The Information Controversy in Black Hole Physics: an Overview}

There are three main criticisms by philosophers of physics that we should consider regarding the use of the term ‘information’ in fundamental physics.

The first criticism concerns an argument that purportedly establishes the Bekenstein-Hawking relation  \citep{bekenstein1973entropy, hawking1976black} between the entropy, $S_{\text{BH}}$, and horizon area, $A_{\text{BH}}$, of a black hole:
\begin{equation}
    S_{\text{BH}}= A_{\text{BH}} / 4 G  ,
    \label{entropyintro}
\end{equation}
in units where $c=\hbar=1$. The most widely accepted view among physicists is that black hole entropy is a genuine thermodynamic entropy. If this is the case, then it is natural to expect that the thermodynamic behaviour of a black hole is underpinned by (quantum) statistical mechanics.

There are essentially two types of arguments supporting this hypothesis in the literature. Assuming that the black hole is made up of a large number of microscopic degrees of freedom, one can either invoke concepts from information theory to relate the entropy to the horizon area, or avoid such concepts altogether. We refer to the former as the `information route' and the latter as the `physical route.' The first route, which has fuelled the information controversy, is described below, while the second will be discussed in Section~\ref{Evidences from Black Hole Thermodynamics}.

The \textit{information route} was developed by Jacob Bekenstein in his influential 1973 article and, as discussed by \cite{wuthrich2017are} and \cite{prunkl2019blackhole}, information-theoretic reasoning plays a central role. Bekenstein interprets $S_{\text{BH}}$ as a measure of the uncertainty an external observer has about the black hole's microstate, given its observed macrostate (assuming that the system possesses microstates: otherwise, knowledge of its macrostate would entail complete certainty). He adopts an \textit{epistemic} understanding of Eq.~\eqref{entropyintro} and, more broadly, subscribes to the view that statistical mechanics concerns our knowledge of the world. 

\citet{dougherty2017blackhole} criticise Bekenstein's interpretation in light of a long-standing debate in the foundations of statistical mechanics. The debate concerns the reduction of thermodynamic entropy, which is undeniably ``mind-independent,'' to (a conception of) Shannon entropy, i.e.~the classical information-theoretic conception of entropy, which changes based on an agent's knowledge about the world, even when the world itself remains unchanged. Applying this discussion to a black hole, the argument is essentially that the facts about its microconstituents should not concern our beliefs or the information we have about them.

Another criticism of the use of the concept of `information' concerns its
connection to communication, in the sense developed in \citet{shannon1948mathematical}'s `A Mathematical
Theory of Communication'. \citet{wuthrich2017are} points out that, before talking about information, a communication system must be in place. This communication system has an inherent minimal complexity, consisting of essentially two elements: an `information source' producing the information and a `communication channel,' through which the information, once suitably encoded, is transmitted. And so, Wüthrich retorts: 
\begin{displayquote}\small
   the complexity of the set-up should not be required of a physical system described by a fundamental physical theory, which is, in general, too impoverished to incarnate all these roles at once at their level \citep[p.~218]{wuthrich2017are}. 
\end{displayquote}
The notion of information being intimately linked to its transmission, Wüthrich asserts that black holes, as described by general relativity, cannot realise the components of the communication system. The conditions for being able to talk about information are therefore not met.

The third criticism, not specifically tailored to black holes, was enunciated by \citet{timpson2013quantum}. Timpson argues against the thesis of ‘informational immaterialism,’ according to which material elements such as particles and fields at the fundamental physical level are replaced by an immaterial basis of information. This view is notably reflected in John Wheeler’s famous slogan ‘‘It from bit'' \citep[pp.~310-311]{wheeler1989}. Timpson contends that the most basic components of the world are not immaterial in nature and that abstractions such as information must be instantiated by something concrete. In his view, we should not regard quantum information theory as introducing a new fundamental material component to the world; some \textit{thing} called `information.'

These objections target particular uses of information: its use in epistemic reasoning to ground black hole thermodynamics, its use where no communication system is available, and its use as a fundamental constituent of reality. They do not show that information-theoretic concepts cannot be used operationally to investigate the informational capacities of a quantum system.

\subsection{The Hayden--Preskill Model and the Way Forward}
\label{The Hayden--Preskill Model and the Way Forward}

Despite this controversy, information remains central to black hole physics. Since \citet{hawking1976breakdown}'s formulation of the black hole information loss paradox, the fate of information in black hole evaporation has been a persistent concern. What role, then, does the concept of information play in contemporary black hole physics?

We address this question by drawing on Timpson's account of the role of quantum information theory in relation to fundamental physics. Timpson argues that quantum information theory need not replace fundamental physical theories or introduce a new immaterial constituent of reality: its value lies in introducing an alternative general framework of enquiry, within which new questions can be asked and devices developed to answer them \citep[p.~239]{timpson2013quantum}. From this perspective, investigating a physical system in terms of its information-carrying, storing, and processing abilities can be a fruitful endeavour.

We argue that the influential work of \citet{hayden2007blackholes} applies quantum information-theoretic reasoning to black holes in precisely this sense. Hayden and Preskill model an evaporating black hole as a quantum channel that transmits quantum information. The first step in their approach, discussed in the remainder of this section, is to treat black holes as ordinary quantum systems. This idea was first proposed by \citet{thooft1985quantum}, further developed by \citet{page1993information}, and has since become so central to the field that it is now referred to as the ‘central dogma’ of black hole physics.  The second step is to take advantage of specific tools from quantum information theory, including the theory of quantum error correction and the concept of entanglement-assisted communication (see~Sects.~\ref{Page’s Model and Scrambling} and \ref{Hayden and Preskill's Model}). In the context of this model, these concepts explain how black holes process new information and how this information is released to the Hawking radiation.

Thus, our slogan `Black holes are about \textit{quantum} information' should not be understood as claiming a form of information immaterialism: it is a \textit{conceptual} and \textit{methodological} slogan that reflects the use of quantum information in the Hayden--Preskill (and cognate) models. We do not mean that black holes \textit{are} quantum information, nor that quantum information fully defines their fundamental nature. This slogan instead highlights how quantum information theory provides tools and a conceptual framework for studying black holes at a fundamental level.\footnote{For more on the philosophical conception and role of information in quantum gravity, see \citet[Section 5]{de2026holography}.}

\subsection{The Central Dogma of Black Hole Physics}\label{The Central Dogma of Black Hole Physics}

The Hayden--Preskill framework requires a physical basis for treating a black hole as an ordinary quantum system. This basis is supplied by the `central dogma' of black hole physics \citep[p.~13]{almheiri2021entropy}. We treat the central dogma not as an established result, but as a modelling assumption specifying the type of system to which quantum information-theoretic tools are being applied.
\newline
\\
\textit{\textbf{Modelling Assumption 0: Central Dogma.} \\ As seen from the outside, a black hole can be described as an ordinary quantum system whose number of available microstates, at specified macroscopic parameters, is determined by its thermodynamic entropy. The full evolution, including its coupling to the environment, is unitary.}

The central dogma states that, up to a chosen cutoff surface beyond which spacetime is treated as fixed, the system composed of the black hole and its atmosphere admits an ordinary quantum description. The black hole’s thermodynamic entropy is given, at leading order in Einstein gravity, by the Bekenstein–Hawking entropy $S_{\text{BH}}$. In our units where $\hbar=c=1$, this can equivalently be written as $ S_{\text{BH}}=A_{\text{BH}}/4\ell_{\text{Pl}}^2$, where  $\ell_{\text{Pl}}^2=G$ is the Planck area. The logarithm of the number of available black hole microstates therefore scales with the horizon area in Planck units. The thermodynamic entropy of the composite black-hole-plus-atmosphere system additionally includes the corresponding exterior-field contribution.

The cutoff surface should not, in general, be identified with the event horizon: rather, it serves as an effective boundary defining the quantum system, whose precise location is framework-dependent.\footnote{For instance, in the context of AdS/CFT \citep{maldacena1999large}, the cutoff surface is placed near the boundary of an AdS spacetime containing a black hole. The gravitational system within this region is then dual to a quantum mechanical system defined on the boundary by imposing a high-energy cutoff on the CFT.} One may thus imagine enclosing the black hole and the surrounding spacetime region up to this surface inside a ``box'' from which signals can be both sent and received. If this hypothesis is correct, an external observer could describe all interactions with the system inside the box within ordinary quantum mechanics.\footnote{This idea appears to be a reformulation of the `postulate 1' of black hole complementarity \citep{susskind1993strechted}, which asserts that a distant observer can describe the formation and evaporation of a black hole entirely within the framework of standard quantum theory. The difference is that the central dogma does not rely on the membrane paradigm or the existence of a stretched horizon, as it makes no commitment to the spatial location of the black hole degrees of freedom \citet[p.~27]{cinti2024beyond}. We will return to this point in Section~\ref{First Model Feature: Number of Qubits}.
}


Up to this point, nothing has been said about the interior structure of the black hole. The central dogma does not capture all the fundamental features that might characterise black holes at a deeper level. In this external picture, the black hole-plus-atmosphere system is portrayed as an ordinary quantum system, with the following properties:\footnote{Other features related to the gravitational nature of the black hole, such as the rapid scrambling of information, will be captured in the Hayden--Preskill framework discussed in Section~\ref{AssumptionsHP}.}
\begin{itemize}

    \item[(a)] It is made up of a finite but large number of degrees of freedom, whose nature is left implicit and which need not be spatially located at the black hole horizon.

    \item[(b)] The statistical behaviour of these degrees of freedom provides a basis for black hole thermodynamics. The full dynamics, including any coupling to the environment, is unitary and generated by a Hamiltonian.
\end{itemize}

We believe that the theoretical evidence supporting the central dogma is robust. It is supported in two ways: first, a number of theoretical developments provide explicit backing for the claim that the Bekenstein-Hawking entropy is a thermodynamic entropy (see~Sect.~\ref{Evidences from Black Hole Thermodynamics}); second, the rejection of certain rebuttals prevents concluding that this claim is simply false (see~Sect.~\ref{Potential Rebuttal from Information Loss}).

\subsubsection{Evidence from Black Hole Thermodynamics}
\label{Evidences from Black Hole Thermodynamics}

The claim that black hole thermodynamics has a statistical-mechanical underpinning is extensively discussed in the philosophy of physics literature.\footnote{Our discussion here will be brief, but the literature is rich; see, for instance, \citet{curiel2014classical,dougherty2017blackhole,wuthrich2017are,wallace2018case,wallace2019case,prunkl2019blackhole}.} It can be summarised in two steps: the first is to establish the thermodynamic behaviour at the level of phenomenological thermodynamics, and the second is to account for it from statistical mechanics (with or without relation to information).

Step one aims to demonstrate that, without going into microscopic considerations, black holes behave like thermodynamic objects.\footnote{\citet{curiel2024Peculiarites} stresses that a thermodynamic-only conception of entropy is necessary for these computations to serve as evidence for black hole statistical mechanics.} \cite{wallace2018case}, whose view we share, insists that Hawking radiation plays a crucial role in transforming the apparent analogy into an identity. Hawking radiation places the black hole in thermal contact with its surroundings, and this interaction occurs in the same way as it would for a thermodynamic system with the corresponding entropy (e.g.~\cite{curiel2014classical,prunkl2019blackhole}).

Step two seeks, in statistical mechanics, as for any thermodynamic system, the fundamental justification for this phenomenological thermodynamic behaviour. While Bekenstein followed the `information route' to statistical mechanics, there exists an alternative perspective. The \textit{physical route} sees the Bekenstein-Hawking entropy as counting the number of microstates available to a black hole with specified macroscopic properties, wherein the entropy changes are grounded in the dynamics of these microscopic degrees of freedom. This is an \textit{ontological} statement about real degrees of freedom realising these microstates and their behaviour. 

This interpretation finds support across several approaches to quantum gravity, which support the Bekenstein-Hawking entropy formula, albeit with varying levels of scope and generality.\footnote{Approaches that reportedly reproduce the Bekenstein-Hawking entropy formula include general relativity treated as a low-energy effective field theory \citep{gibbons1977action,gibbons1978path}, string theory \citep{strominger1996microscopic}, loop quantum gravity \citep{ashtekar1998quantum,bianchi2012entropy}, and the AdS/CFT correspondence \citep{maldacena1999large, strominger1998blackhole}. See \cite{wallace2019case} for more details.} For \citeauthor{wallace2019case} (\citeyear{wallace2019case}, p.~30), it is not remotely plausible that all of this is a massive coincidence.\footnote{Wald's Noether-charge construction shows that, for a general diffeomorphism-covariant gravitational theory, the entropy of a stationary black hole is fixed by the gravitational action; for Einstein gravity, it reduces to the Bekenstein-Hawking formula \citep{wald1993blackhole}. Thus, agreement with the area law is perhaps not completely unexpected.} He argues that making sense of this convergence requires viewing these approaches as partial descriptions of a consistent quantum theory of gravity (p.~2).

Returning to the status of `information,' which is at stake in the present paper, \citeauthor{wallace2018case} (\citeyear{wallace2018case}, p.~3) further emphasises that Bekenstein’s informational motivation is no longer necessary: modern developments provide a strong foundation for black hole thermodynamics in statistical mechanics, \textit{without any further appeal to `information.'}

The view we defend in this paper is that these developments do not warrant discarding the notion of information altogether. These computations support the claim that black hole entropy has a microscopic statistical basis, without establishing a particular microscopic ontology. This motivates exploring a finite-dimensional quantum model of the black hole as a modelling hypothesis and paves the way for an abstract characterisation of black holes through information-theoretic reasoning. 

\subsection{Possible Rebuttal of the Central Dogma from a Paradox}
\label{Potential Rebuttal from Information Loss}

Our next concern is the black hole’s dynamics. To satisfy the central dogma, its evolution as seen from the outside must be {\it unitary}. However, anyone familiar with \citet{hawking1976breakdown}'s ‘information loss paradox’ will burn to object that the complete evaporation of a black hole is a process that violates unitarity.  

Inconsistencies between Hawking's account of evaporation and the external picture given by the central dogma appear, in fact, well before the end of the evaporation. The realisation of these premature troubles stems from the `Page time paradox,' introduced by \citet{page1993information}. It is \textit{this} version of the information loss paradox that, as articulated by \citet{wallace2017why}, practising physicists regard as genuinely paradoxical.

More recently, work based on the path integral approach to gravity has produced results compatible with unitary black hole evaporation, thereby keeping open the possibility that the central dogma holds.

\subsubsection{Page time Paradox}
\label{Page Time Paradox}

\citet{page1993information}'s idea was to incorporate evaporation into the simple picture of a black hole given by the central dogma.  Following \citet{wallace2017why}, assuming the black hole starts in a pure state, unitarity requires that the combined system of black hole plus radiation remains pure. Page demonstrated that, if the central dogma is true and the black hole is a quantum statistical mechanical object, respecting unitarity results in a bound, imposed by the Bekenstein-Hawking entropy, on the number of Hawking modes entangled with the black hole degrees of freedom. The so-called `Page bound' reads:
\begin{equation}
   S_{\text{vN}}(\rho_{\text{Rad}}(t)) \leq  S_{\text{BH}}(t)\,,
   \label{Pagebound}
\end{equation}
where $S_{\text{vN}}(\rho_{\text{Rad}}(t))=- \operatorname{tr}[{\rho}_{\text{Rad}}(t) \log {\rho}_{\text{Rad}}(t)]$ is the von Neumann entropy of the radiation and $S_{\text{BH}}(t)= \log \operatorname{dim} \mathcal{H}_{\text{BH}}(t)$ is the black hole entropy.

The {\it Page time paradox} is as follows. As the black hole evaporates, $S_{\text{vN}}(\rho_{\text{Rad}}(t))$ of the emitted radiation increases through their entanglement with degrees of freedom behind the horizon \citep{hawking1975particle}, while $S_{\text{BH}}(t)$ decreases as the black hole area diminishes. The time at which the Page bound saturates is known as the `Page time' (roughly half of the evaporation time for a Schwarzschild black hole). The radiation entropy cannot continue to follow Hawking's increasing curve beyond this point as there are no longer enough black hole modes to maintain the thermal nature of the radiation.

There are at least two possible conclusions to draw from this violation of the Page bound \eqref{Pagebound}, neither of which is particularly attractive given that both the Page bound and Hawking's calculation are expected to be reliable: 

(a) One can abandon the central dogma, and with it the explanation for the black hole's thermodynamic behaviour. Black holes would then be unique thermodynamic systems without underlying statistical mechanics. One would also need an alternative, non-miraculous explanation for the derivation of black hole entropy in the different quantum gravity approaches discussed above. 

(b) One can call the result of Hawking’s calculations into doubt and argue that Hawking radiation has non-thermal components. This is the option taken by Page in his paper, to which we return in Section \ref{Page's Modelisation and Results}. If true, then the apparent breakdown of QFT in a curved background, i.e.~its incorrect prediction of a loss of unitarity, must be explained, especially since it occurs well within the theory’s expected domain of applicability.

\subsubsection{A Potential Resolution of the Page Time Paradox}
\label{PotentialRes}

Recent works propose a resolution consistent with the central dogma, thus favouring option (b) above.\footnote{\citet{almheiri2021entropy} provides a review of these findings tailored for readers with minimal prior knowledge of the subject and includes the relevant references.} These works derive a gravitational formula for computing the von Neumann entropy of quantum systems coupled to gravity \citep{lewkowycz2013generalised,faulkner2013quantum,dong2016deriving,dong2018entropy}. 

This proposal was concretely implemented for evaporating black holes in AdS/CFT \citep{almheiri2019entropy,almheiri2020page,penington2020entanglement}, and subsequently in two-dimensional gravitational models without appeal to holography, by incorporating `replica wormhole saddles' into the gravitational path integral \citep{penington2022replica,almheiri2020replica}.\footnote{See \citet{cinti2024beyond} for a more detailed philosophical analysis.} In both cases, the resulting entropy of Hawking radiation respects the Page bound throughout evaporation.

These arguments rely on idealisations, such as working in low spacetime dimensions or in AdS, so there is not yet a fully definitive solution for an ordinary Schwarzschild black hole. Nonetheless, this suggests that the semiclassical prediction of a continually increasing radiation entropy need not survive a more complete gravitational treatment, and that the paradox for the Schwarzschild case may be resolved along similar lines. If so, the central dogma may well remain a robust statistical mechanical foundation for explaining black hole thermodynamics.

\section{The Dawn of Quantum Circuits: Page’s Model}\label{Page’s Model and Scrambling}

Quantum circuits provide models of the time evolution of quantum systems; according to the central dogma, this makes them natural models for the dynamics of black holes. The relevant quantum information is encoded in the initial state of (a subset of) qubits, which evolves under the system’s dynamics. In this way, the black hole performs a computation in which information is processed but not destroyed.

Even though quantum circuit models presuppose unitary evolution, they remain a valuable framework for investigating its consequences. As discussed in the previous section, \citet{page1993information} also assumed unitary evolution and thereby derived a conclusion that appears to conflict with the semiclassical quantum-field-theoretic description of Hawking radiation. In this section, we argue that Page went beyond identifying the Page time paradox: his work can be interpreted as pioneering the use of quantum circuit models of black holes to show how unitary evaporation could yield an entropy curve compatible with the Page bound.

We also discuss the notion of `scrambling.' Its use in the context of black hole evaporation can be traced retrospectively to \citet{page1993average}, and it is now regarded as central in black hole physics. In particular, scrambling is a cornerstone of the later and more refined model proposed by Hayden and Preskill.

Section \ref{Quantum Circuit Model} provides a general introduction to quantum circuits and outlines the key features that specify any quantum circuit model of black holes. Section \ref{Pages Model in Relation to Scrambling} delves into the details of Page's quantum circuit model, where the concept of scrambling first emerges.

\subsection{Quantum Circuit Model of Black Holes}
\label{Quantum Circuit Model}
 
In Section \ref{Introduction to Quantum Circuits}, we introduce the basic components of quantum circuits and what they are designed to represent. We then turn, in Section \ref{Quantum Circuit-Model of Black holes}, to the key features required when using these models to describe how black holes process information.

\subsubsection{Introduction to Quantum Circuits}
\label{Introduction to Quantum Circuits}

Quantum circuits are visual and algebraic representations of quantum dynamics, in which a global evolution is approximated by a finite sequence of simple operations on a finite set of subsystems. By `simple operation,' we mean a unitary operator, or gate, acting on a small number of qubits, typically two. Such gates are usually selected from a universal gate set, i.e.~a relatively small collection of gates from which finite sequences can approximate any unitary operation to arbitrary accuracy.

The \textit{all-to-all} quantum circuits considered here consist of sequences of such gates in which any pair of qubits can, in principle, interact. A given circuit represents the construction of a particular unitary operator $U$ by applying a sequence of gates from the universal set $\{g_1, g_2, \dots, g_n\}$:
\begin{equation}
    U = g_i g_{i-1} \ldots g_1 I.
    \label{Operatorasgate}
\end{equation}
Equivalently, it models how an output state is obtained from an input state by the action of this unitary.

The all-to-all character of a circuit is a component of its `architecture': the pattern of possible interactions between its qubits. In this case, interactions are not limited by spatial locality or fixed neighbourhood relations. A circuit can also exhibit `parallel processing' when, at each step, several gates act simultaneously on disjoint pairs of qubits.

\begin{figure} 
    \centering
    \includegraphics[scale=0.65]{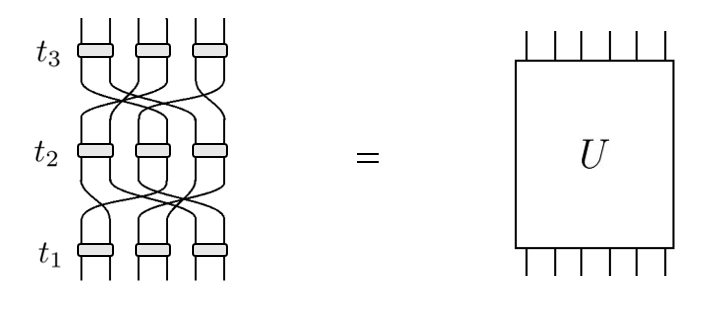}
    \caption{\small All-to-all circuit with parallel processing. At each step, every qubit interacts with one other qubit by means of a gate, although no qubit interacts with more than one other qubit at that step. The pairings can change between steps. The circuit depicts the gradual transformation of a state of six qubits (the lines) by the action of gates (the boxes).}
    \label{2localcircuit}
\end{figure}

One can equip circuits that exhibit parallel processing with a clock. The resulting notion of time is called `circuit time' and is denoted by $t_n$, where $n$ indexes the circuit steps. In the diagram, circuit time runs in the vertical direction. The circuit's `depth' is the number of such time steps, while its `size' is the total number of gates.

Finally, although quantum circuits can in principle approximate arbitrary unitary operations, doing so may require a circuit of exponential size.

\subsubsection{Modelling Black Holes: Features of the Model}
\label{Quantum Circuit-Model of Black holes}

Let us now discuss the specific features of quantum circuits as models of black holes. As we now explain, the significance of such a model is that its specification requires modelling choices. These choices make explicit physical assumptions that are otherwise hidden in claims that `a black hole processes information.'

A quantum circuit features, at the very least, a finite number of qubits and a sequence of gates acting on them. These components are precisely what the central dogma gives us (see~Sect.~\ref{The Central Dogma of Black Hole Physics}): property (a) stipulates that the black hole is constituted by a finite number of degrees of freedom; these subsystems can be treated as the qubits on which the gates act. Property (b) indicates that the dynamics is unitary and can, in principle, be approximated to arbitrary accuracy by a quantum circuit.

That being said, additional modelling assumptions and idealisations remain to be made in order to determine the exact characteristics of the model and render it applicable to problems in black hole physics. We identify three key features that must be specified in order to define a quantum circuit model (QC-model) of a black hole. We will call these the {\it features of the model.}
Each of these features reflects details of the physical object it represents, namely the black hole:
\begin{enumerate}
 \item[(1)] the {\it number of qubits} determines the dimension of the Hilbert space characterising the black hole, 
 \item[(2)] the {\it circuit architecture} and {\it size} represent the type of Hamiltonian generating the internal dynamics, and the efficiency with which the latter can be approximated, 
    \item[(3)] the {\it initialisation of the qubits} is the state in which the black hole starts.
    
\end{enumerate}

To fully specify the circuit model, choices must be made for each of these features. Although informed by what is known and believed about black holes, these choices remain hypotheses in the absence of an observationally confirmed theory of quantum gravity. Moreover, as is to be expected of a model, the representation of the black hole it provides is simplified, and its construction requires idealisations. Most importantly, relying on the central dogma, it adopts an external perspective and does not represent the detailed physics near or behind the event horizon. It also represents information processing at a highly abstract level, independently of the detailed physical interactions between the particles that instantiate it.


The specific assumptions and idealisations underlying the black hole model proposed by Hayden and Preskill will be laid out in Section \ref{AssumptionsHP}.

\subsection{Page's Model in Relation to Scrambling}
\label{Pages Model in Relation to Scrambling}

Section \ref{Scrambling: Definition and Relevance for Black Holes} introduces the concepts of `Haar scrambling' and the weaker notion of `Page scrambling,' both of which originate from Page's approach. Section \ref{Page's Modelisation and Results} then presents Page's analysis as a quantum circuit model of a black hole and highlights the connection between his results and different notions of scrambling.

\subsubsection{Scrambling: Definitions and Distinctions}
\label{Scrambling: Definition and Relevance for Black Holes}

The term ‘scrambling’ is sometimes used to refer to a property of a quantum state, namely its entanglement structure, or sometimes to a property of unitary dynamics, namely its ability to delocalise quantum information \citep{sekino2008fast,hosur2016chaos}. There are multiple, non-equivalent notions of scrambling
already at the state-level, so some caution is required.

The strongest notion of scrambling is based on a
`Haar-random unitary:' a unitary transformation selected uniformly at random
according to the Haar measure on the relevant unitary group, i.e.~$U(2^b)$ for a system of $b$ qubits, and which serves as the standard reference for fully generic quantum dynamics.\footnote{The Haar
measure properly weighs different regions of the unitary group, allowing one
to sample its elements uniformly given the structure of the group. For the
mathematical foundations of the Haar measure, see
\citet{duistermaat1999lie}.} Applying such a unitary to a fixed reference
state generates a typical pure state. This leads to the following definition of a ‘Haar-scrambled’ system \citep{sekino2008fast}:
\begin{itemize} \item[\textbf{}] A system is \textbf{Haar-scrambled} when its quantum state has been randomised with respect to the Haar measure over the entire Hilbert space.
\end{itemize}

With overwhelming probability, such random states have near-maximal
entanglement \citep{hayden2006aspects}. This motivates the definition of a weaker property for the entanglement structure of a state:
\begin{itemize}
    \item[\textbf{}] A system in a pure state is \textbf{Page-scrambled} when every subsystem smaller than half of the whole system has entanglement entropy close to its maximal value.
\end{itemize}
The closeness refers to the entropy deficit in the sense specified below. Haar-random states are typically Page-scrambled, as \citet{page1993average} showed and as we review below, and obtaining a Page-scrambled state does not require that the dynamics be exactly Haar-random.

This distinction will be important for understanding the difference between Page’s model and that of Hayden and Preskill, discussed in Section~\ref{AssumptionsHP}. It should be noted at this stage that implementing a generic Haar-random unitary requires a circuit of exponential size in the number of qubits, and this limitation will be of particular concern for the protocol of Hayden and Preskill.

\subsubsection{From Typical States to Quantum Circuits} 
\label{Page's Modelisation and Results}

\cite{page1993information} proposed a model of an evaporating black hole in which its emitted radiation and the remaining black hole are treated as complementary subsystems of a closed system in a \textit{typical} pure state and the size of the black hole subsystem  is tied to the Bekenstein-Hawking entropy.

This random-state model can be represented, retrospectively, by a quantum circuit: one begins with a fixed reference state of a total $b$-qubit system, applies a unitary chosen uniformly with respect to the Haar measure on $U(2^b)$, and then partitions the resulting state into a $d$-qubit subsystem $D$ identified as the radiation and the complementary $(b-d)$ qubits as the black hole subsystem. In this circuit representation, the black hole's dynamics is modelled  by a Haar-random unitary; Page's original argument, however, concerns the typical entanglement structure of the resulting state rather than a detailed microscopic dynamics.

Page then used this model to study how the entanglement between the black
hole and its radiation depends on the relative sizes of the two subsystems. He showed that the entanglement entropy of the radiation subsystem is close to maximal in a typical Haar-scrambled pure state, so long as it is smaller than half the size of the whole system. Quantitatively, 
\begin{equation}
    S_{\text{vN}}(\rho_D)=d-O\left(e^{2d-b}\right)
    \label{Pageresult}
\end{equation}
for all $d<b / 2$, which means that the black hole-plus-radiation system is Page-scrambled with overwhelming probability 
\citep[p.~1292]{page1993average}. This equation expresses that the entanglement between the radiation and the black hole starts small and increases until half the black hole entropy has been radiated.

In relation to information loss (or, more accurately, information release), the intuition behind Page's result is that information is encoded non-locally in correlations between the black hole and its radiation. Before the Page
time, the radiation subsystem is too small to
reveal information about the initial black hole state. After the Page time, information emerges through increasingly significant deviations from exact thermality of the radiation state. Later radiation can purify the early radiation, preserving the purity of the black hole-plus-radiation system throughout.

Thus, within this model, combining Haar-random dynamics with evaporation ensures that the Page bound \eqref{Pagebound} is respected throughout evaporation. This result underpins option (b) of Section~\ref{Page Time Paradox} for resolving the Page time paradox.

\section{Communication Perspective: the Hayden--Preskill Model}
\label{Hayden and Preskill's Model}

Page's analysis, discussed in the previous section, does not explain why scrambling is crucial for information to escape, nor what conditions an observer must satisfy to recover it.
To further investigate these aspects of the Page time paradox, \citet{hayden2007blackholes} extended Page’s QC-model by introducing additional modelling assumptions that allow a black hole to participate in a communication protocol.

Their proposal is formulated as a thought experiment involving two protagonists, Alice and Bob. Alice encodes her secrets in a quantum diary (e.g.~the memory of a quantum computer) which she throws into a black hole. The black hole evaporates, loses mass and eventually vanishes, encoding Alice’s diary in the outgoing Hawking radiation where it might be decrypted by Bob. Since the black hole evaporates extremely slowly (with a time scale of order $M_{\text{BH}}^3$ in Planck units), Alice believes that her secrets will be safeguarded for many generations. Bob wants to access her confidences from Hawking radiation before the black hole fully disappears.

Section \ref{AssumptionsHP} outlines the assumptions defining their QC-model, their physical significance, and how these differ from Page's. Section \ref{Details on the Procedure for Information Release} then explains how this model represents the black hole as a quantum channel relying on entanglement to protect (see~Sect.~\ref{Entanglement-Protected Communication}) and assist (see~Sect.~\ref{Entanglement-Assisted Communication}) the transmission of information. Section \ref{Questions Tackled and their Answers} gives the model's answer to two quantitative questions about the retrieval of quantum information from the black hole, mentioned in the Introduction. Finally, Section \ref{Information Without Controversy} addresses how Hayden and Preskill’s approach avoids earlier criticisms about the use of information in black hole physics.

\subsection{Specifying the Model: Assumptions and Idealisations}
\label{AssumptionsHP}

In this section, we outline the anatomy of the Hayden–Preskill QC-model of a black hole in terms of its three main features (see~Sect.~\ref{Quantum Circuit-Model of Black holes}). Fully specifying this model requires eight assumptions, on top of the central dogma. We discuss their theoretical basis, some of which was developed after the publication of Hayden and Preskill’s work, and place them in the broader context of the existing literature. Our aim is to distinguish between the types of assumptions underpinning their decoding result.

\subsubsection{First Model Feature: Number of Qubits}
\label{First Model Feature: Number of Qubits}

The central dogma implies that, from the outside, a black hole can be modelled as a quantum system $B$ with effective Hilbert-space dimension determined by its thermodynamic entropy. Representing this system as $b$ qubits, this means that the overall Hilbert space is given by the tensor product: 
\begin{equation}
\mathcal{H}_{B}=\mathcal{H}_{1} \otimes \cdots \otimes \mathcal{H}_{b}\,,
\end{equation}
where the $\mathcal{H}_{i}$'s are the single-qubit Hilbert spaces.

The spatial location of these degrees of freedom is not provided by the central dogma. Hayden and Preskill assume that the $b$ degrees of freedom are uniformly distributed over the black hole’s `stretched horizon,' a timelike surface or `membrane' extending around the event horizon of the black hole at a Planck-length proper distance from it.

\begin{definition} \textbf{Location of the Degrees of Freedom.} \\
The black hole degrees of freedom are uniformly distributed on the stretched horizon, with about one degree of freedom per Planck unit of area. 
\label{Locationdof}
\end{definition}
For them, the stretched horizon also serves as the cutoff separating the system $B$ from its external environment. This specifies the first feature of the QC-model of this black hole: there are $b\approx A_{\text{BH}} / 4G$ qubits, where $A_{\text{BH}}$ is the area of the event horizon at the Schwarzschild radius $r_S$ and where we suppressed the conventional factor of $\operatorname{ln}2$. 

This assumption reflects Hayden and Preskill's subscription to the `membrane paradigm' \citep{damour1979quelques, thorne1986membrane,susskind1993strechted}; see \citet{wallace2018case, wallace2019case} for the quantum extension. The paradigm is based on the premise that the stretched horizon could legitimately be considered as the repository of the quantum information absorbed by the black hole, providing a statistical-mechanical underpinning for the black hole’s thermodynamics.\footnote{
Unlike the event horizon, originally defined globally by \citet{hawkingellis1973large,wald1984general} as a region \textit{of spacetime} with implicit reference to all future times \citep[p.~9]{Curiel:2018cbt}, the stretched horizon can be treated as a timelike membrane enclosing a region of space, evolving through time, and carrying time-dependent thermodynamic properties \citep[p.~18]{wallace2019case}.}

\citet{cinti2024beyond} have qualified this choice of spatial location as \textit{conventional} because it does not reflect any fundamental feature of the full quantum gravity theory. They explain that `the idea of a smooth membrane exactly located one Planck length from the horizon is exactly the sort of geometric notion one expects to not make sense in quantum gravity in which spacetime structure is not well-defined' (p.~27).  The semiclassical description of the black hole exterior nevertheless remains valid, and the stretched horizon should therefore be understood as an effective construct. Moreover, even from a classical general relativistic viewpoint, the status of the membrane paradigm is not entirely uncontroversial.\footnote{For instance, \citet{padmanabhan2011entropy} derives the thermodynamic behaviour of null surfaces directly, thereby avoiding the need for a heuristic membrane description.}

With recent developments leading to the formulation of the central dogma, locating the degrees of freedom that describe a black hole on the stretched horizon is no longer the only possibility considered by high energy physicists. For instance, in holographic approaches, these degrees of freedom are instead thought to reside on the boundary of spacetime. Another example comes from string theory, where the counted microstates correspond to configurations of D-branes, extended objects that are not understood to be localised at the horizon, the very notion of a classical horizon not being manifest at this level of description.

\subsubsection{Second Model Feature: Circuit Architecture and Size}
\label{Second Model Feature: Circuit Architecture and Size}

The central dogma further stipulates that the evolution of a black hole’s microscopic degrees of freedom, from the viewpoint of an external observer, is governed by a Hamiltonian whose exact form is not specified. The gravitational description, in contrast, is subject to a Hamiltonian constraint associated with diffeomorphism invariance, while boundary terms can generate evolution with respect to an asymptotic time. These structures do not, however, directly specify the microscopic Hamiltonian posited by the central dogma, which is therefore not manifest in the classical description.

While extracting a Hamiltonian that describes only the evolution accessible to an external observer might be nontrivial, if it exists, it must reproduce the expected spacetime dynamics  \citep[p.~14]{almheiri2021entropy}. We now discuss how various shockwave computations lead to the expectation that this Hamiltonian generates dynamics characteristic of quantum chaos and how this is implemented in the present QC-model.

As a starting point, Hayden and Preskill assume that interactions between all the degrees of freedom are allowed, but only a few at a time—making the interactions local in this sense.

\begin{definition} \textbf{Easy Dynamics.} \\
The internal dynamics of the black hole is governed by an `easy Hamiltonian,' i.e.~a sum of simple Hermitian operators involving interactions between all degrees of freedom, but coupling only two of them at a time.  
\label{Easy Dynamics}
\end{definition}

This modelling assumption is an idealisation of the black hole's dynamics, making it suitable for modelling with an all-to-all quantum circuit of two-qubit gates with parallel processing (see Sect.~\ref{Introduction to Quantum Circuits}), whose evolution is analogous to that generated by such an easy Hamiltonian \citep[p.~6]{susskind2016entanglement}.

In spite of the fact that the black hole's dynamics is modelled as being ``easy'', it is assumed that it evolves according to a Haar-random unitary, introducing a degree of ``complexity'' through scrambling:

\begin{definition} \textbf{Scrambling Dynamics.} \\
 The black hole's dynamics conceals the diary information by scrambling it, i.e.~thoroughly mixing it among all the black hole degrees of freedom in a highly non-local form. 
\label{Scrambling Dynamics}
\end{definition}
With this assumption, the choices and hypotheses in the Hayden--Preskill QC-model start to deviate from those made by \citet{page1993information}.

Page explained how information about the initial black hole state could not be revealed from sufficiently small subsystems due to correlations between the black hole and its radiation (see Sect.~\ref{Pages Model in Relation to Scrambling}). On the other hand, Hayden and Preskill are interested in the delocalisation of initially \textit{local} quantum information. In the QC-model, this dynamical delocalisation is idealised by taking the evolution to be Haar-random. Through Haar scrambling, information about a small part of the input (the diary) becomes inaccessible from parts of the output that are too small. This scrambling is not simply a property of the output state, but also a dynamical notion: the two being, in fact, closely related \citep{hosur2016chaos}.\footnote{The delocalisation of local information can be analysed in terms of the entanglement on a Hilbert space double the size of the input, a property that will be reflected in the choices made for the initialisation of the qubits discussed in Section~\ref{Third Model Feature: Initialisation of the Qubits} and explained in Section~\ref{Entanglement-Protected Communication}.}

This delocalisation is closely related to the growth of operators: under chaotic dynamics, initially simple operators evolve into increasingly complex ones that act non-trivially on larger and larger subsets of the system \citep{shenker2014blackholes, kitaev2015hidden}. As a result, an operator that initially perturbs only a few degrees of freedom becomes spread across the entire system over time.

The scrambling of quantum information in this way can be diagnosed by the decay of `out-of-time-order correlation functions,' i.e.~four-point correlation functions that measure the failure of commutation between operators at different times. Their decay is a signature of the chaotic nature of the system, in which a local perturbation can have widespread effects over time, in a manner akin to the classical notion of sensitivity to initial conditions.\footnote{\citet{maldacena2016bound} have conjectured a bound on the influence of chaos on such correlators: $\left\langle O_A(0) O_B(t) O_C(0) O_D(t)\right\rangle\sim1- \exp\{\lambda t\}/n$, where $O(t)=U^{\dagger}(t) O U(t)$, $n$ is the number of degrees of freedom in the system and the Lyapunov exponent
$\lambda$ is bounded by $\lambda\leq {2 \pi}T$.}

Another difference with Page's approach is that Hayden and Preskill demand that the information deposited by Alice is \textit{rapidly} scrambled by the black hole’s dynamics, `on a time scale comparable to the time interval between the
emission of successive radiation quanta' (p.~10). This is characterised by an important time scale:
\begin{itemize} \item[\textbf{}] The \textbf{scrambling time} $t_{\mathrm{scr}}$ is the time needed for a Page-scrambled system to re-scramble when one qubit of information is added.
\end{itemize} 
Thus, Hayden and Preskill propose that:
\begin{definition} \textbf{Rapid Scrambling Constraint.} \\
The scrambling time is \textit{rapid}, i.e.~it occurs on a time scale comparable (up to a logarithmic factor) to the time interval $r_S$ between the emission of successive radiation quanta, as measured by a distant observer.\footnote{To determine the time interval between successive radiation quanta, one can divide the black hole's evaporation time, $M_{\text{BH}}^3$, by the total number of emitted qubits, $M_{\text{BH}}^2$, and use the relation $r_S\sim M_{\text{BH}}$ to express the result in terms of the Schwarzschild radius.}
\label{Rapid Scrambling Constraint}
\end{definition}
This bound poses a physical constraint on the scrambling time and ensures that the radiation escaping the black hole can contain recoverable information about Alice's quantum state.

How is the rapid scrambling constraint expressed in the QC-model? Addressing this question requires a slightly more technical discussion. As mentioned earlier, generating the action of a Haar-random unitary on an input state requires a quantum circuit of exponential size in the number of qubits \citep[Sect.~4.5.4]{nielsen2010quantum}. Consequently, if the evolution operator of the black hole is described by a Haar-random unitary, the quantum circuit that models its internal dynamics has to be of exponential size. Not only is this unrealistic for physical evolution, but this also prevents information from being mixed quickly enough to satisfy the rapid scrambling constraint.

The way \citet{hayden2007blackholes} find to accommodate \textit{Modelling Assumption \ref{Scrambling Dynamics}} with \textit{Modelling Assumption \ref{Rapid Scrambling Constraint}} is to rely on a \textit{weaker} notion of Haar scrambling in their protocol: the unitary $U$ is sampled from an ensemble that approximates the Haar measure for the corresponding analogue of the task they initially considered. This is reflected in the following hypothesis:

\begin{definition} \textbf{Approximate Decoding.} \\
Bob is required to recover Alice's diary only approximately, with a small reconstruction error that determines the precision of the decoding task.
\label{Approximate Decoding}
\end{definition}
In other words, Bob only needs to recover from the radiation a quantum state that is \textit{close}, according to some suitable measure of similarity, to Alice's original state (see~Sect.~\ref{Entanglement-Assisted Communication}). For this decoding task, the unitary $U$ need not be sampled from the full Haar ensemble. This relaxation allows the encoding process to be implemented using simpler and more efficient circuits.

Specifically, this allows the results to hold for a carefully chosen subset of operations that behave like random ones for the purposes of this communication protocol. The unitary $U$ is sampled from a subset of the full unitary group known as the `$\epsilon$-approximate unitary 2-design' \citep{dankert2009exact} and can be generated by gate sequences of smaller size. In particular, generating such unitaries requires a circuit of depth approximately $\log (b) \log (\epsilon^{-1})$ (neglecting the diary size compared to the number of black hole qubits), where $b$ is the number of black hole qubits and $\epsilon$ controls the quality of the approximation.

Finally, if one further requires the circuit gates to respect the stretched horizon’s spatial geometry, there is an additional cost in the circuit depth from the time it takes for the information to propagate from one location to the other. This cost is determined by the linear size of the system. As the qubits are uniformly distributed over the stretched horizon, with a radius of order $r_S\sim \sqrt{b}$, the additional cost of using geometrically-local gates is, at worst, a multiplicative factor of the order of $\sqrt{b}$. Therefore, the circuit scrambling time estimate becomes
\begin{equation}
    t_{n,\mathrm{scr}} \approx \sqrt{b} \log (\sqrt{b}) \log (\epsilon^{-1}),
    \label{tscramblingcircuit}
\end{equation}
where we neglected a factor of $2$.

Of course, a black hole does not evolve through a (quasi) Haar-random unitary selected—by whom?—at random from all possible unitaries in the group. Is there another physically meaningful mechanism by which black holes scramble information, and does it operate quickly enough to satisfy the rapid scrambling constraint?

The black hole scrambling time was first estimated by \citeauthor{hayden2007blackholes} (\citeyear{hayden2007blackholes}, Sect.~4). This estimate is based on an essentially classical argument concerning the exponentially rapid diffusion of a charged particle dropped on the stretched horizon, originally derived by \citet{thorne1986membrane}. A charged particle dropped into a black hole spreads rapidly on the stretched horizon, with the proper area of a droplet of charge growing like $\exp \left(t_S / r_S\right)$. Assuming that this exponential spreading also takes place when a \textit{piece of quantum information} is deposited on the horizon, they obtained a Schwarzschild scrambling time estimate of the order $r_S \log (r_S/\ell_{\text{Pl}})$. This classical picture was later extended by \citeauthor{sekino2008fast} (\citeyear{sekino2008fast}), who elaborated on the estimate in the context of string theory, arriving at a similar scrambling timescale.

Modern understanding of black hole scrambling, on the other hand, emphasises the role of chaotic time evolution and can be diagnosed by the decay of out-of-time-order correlation functions. In gravitational systems, these correlation functions can be explicitly computed using gravitational shockwaves. When a weak perturbation falls into a black hole, it experiences an exponential blueshift and creates a shockwave near the horizon that disrupts later infalling matter. This effect, first studied for Schwarzschild black holes by \citet{dray1985effect}, was later connected to quantum chaos by \citet{kitaev2015hidden}, who showed that such gravitational dynamics effectively calculate out-of-time-order correlation functions in disguise.\footnote{Extensions include AdS \citep{shenker2014blackholes}, Reissner–Nordström \citep{sfetsos1995gravitational}, and Kerr black holes \citep{Bentov2019gravitational}.} The shockwave causes the out-of-time-order correlation functions to decay, an effect that signals the onset of scrambling and becomes significant at timescales of order $r_S \log (r_S/\ell_{\text{Pl}})$.

Hence, there is substantial theoretical evidence for a black hole scrambling timescale of order
\begin{equation}
        t_{\mathrm{scr}} \approx  r_S \log (r_S/\ell_{\text{Pl}}),
        \label{tscrambling}
    \end{equation}
which satisfies the rapid scrambling constraint. This timescale arises not from Haar-random unitary dynamics, but from the chaotic evolution governed by gravitational interactions near the horizon.

We can calibrate the circuit time against the continuous time evolution by writing $t_n=t/\Delta t $, where $\Delta t$ is the asymptotic time interval between successive circuit steps. Using $b \approx(r_S/\ell_{\text{Pl}})^2$ and comparing Eqs.~\eqref{tscramblingcircuit} and \eqref{tscrambling}, we obtain $\Delta t=t_{\text{Pl}}$, up to numerical factors and for fixed $\epsilon$. Thus, from the perspective of distant observers, each layer of the geometrically local circuit corresponds to an interval of approximately one Planck time.

A potentially problematic consequence of this modelling choice is the following. As Hayden and Preskill note, gravitational redshift makes a Planck-time interval in Schwarzschild time correspond to a much shorter proper-time interval for static observers near the stretched horizon \cite[p.~11]{hayden2007blackholes}.
If the circuit is interpreted literally as describing local dynamics at the stretched horizon, each gate would therefore operate on sub-Planckian timescales.

This may sound similar to the ‘trans-Planckian problem’ associated with Hawking radiation \citep{jacobson1991blackhole}. Although Hawking radiation is detected far from the black hole, at relatively long wavelengths and late times, its outgoing modes are sensitive, when traced backwards, to physics near the event horizon at ultra-short wavelengths. The concern is therefore that quantum spacetime effects may be amplified so as to call into question the semiclassical framework used by \cite{hawking1975particle} to model black holes. However, we do not view the present issue in quite the same light. The description in terms of a quantum circuit is intended solely as an effective representation of the black hole's dynamics from the perspective of distant observers; its gates need not correspond to local microscopic processes on the stretched horizon.

\subsubsection{Third Model Feature: Initialisation of the Qubits}
\label{Third Model Feature: Initialisation of the Qubits}
The third feature of the model concerns the initial state of the components in the thought experiment, which consists of two maximally entangled bipartite states, each written in EPR form.

The first maximally entangled state is introduced by coupling the diary qubits to an external reference system. One assumes that Alice’s quantum diary is maximally entangled with a reference system, and the qubits of $A$ are then thrown into the black hole:
\begin{definition} \textbf{Reference System.} \\
The quantum diary $A$ is maximally entangled with a reference system $R$ of the same dimensionality, i.e.~their initial joint state is the pure state:
\begin{equation}
    |\mathrm{EPR}\rangle_{R A}= \frac{1}{\sqrt{ 2^a}} \sum_i|i\rangle_R \otimes|i\rangle_A\,,
\end{equation}
where $\mathcal{H}_R \equiv \mathcal{H}_A$ and $\{|i\rangle\}$ denotes the standard basis of $\mathcal{H}_A$. 
\label{Reference System}
\end{definition} 
The reference system does not necessarily represent a physical system. Its introduction is a mathematical trick that allows us to analyse the decoding of all possible diary states simultaneously, rather than that of a specific quantum state. This, in turn, enables a quantitative discussion, in terms of information-theoretic measures, of when the information is successfully transmitted.
The key insight is that successful transmission to the radiation implies that the information is, in principle, recoverable and that a decoder exists. The existence of such a decoder can be diagnosed by the amount of entanglement between the reference system and the radiation (see~Sect.~\ref{Details on the Procedure for Information Release}).

The next assumption builds on an observation due to \citet{page1993information}.
\begin{definition} \textbf{Old Black Hole.} \\
The black hole, which was formed in a pure state, has been evaporating for a time longer than the Page time.
\label{Old Black Hole}
\end{definition}
This modelling assumption specifies both the system under consideration (namely, an old black hole) and an initial condition for the black hole (namely, that its initial state is pure). After the Page time, the Hilbert space of the radiation is larger than that of the black hole, and the remaining black hole is nearly maximally entangled with a suitable subsystem of the early radiation (see~Sect.~\ref{Page's Modelisation and Results}).

The central insight is that a distant observer with access to all the emitted radiation could then probe more than half the system and detect non-thermal features. 
Hayden and Preskill assume the existence of such an observer, introducing a second maximally entangled state:
\begin{definition} \textbf{Bob's Unlimited Control.} \\
By collecting all the Hawking
radiation emitted by the old black hole, Bob holds a quantum memory, within which a subsystem $B'$ is maximally entangled with the state of the black hole $B$ in the EPR state
\begin{equation}
    |\mathrm{EPR}\rangle_{BB'}= \frac{1}{\sqrt{ 2^b}} \sum_i|i\rangle_B \otimes|i\rangle_{B'}\,,
\end{equation}
where $\mathcal{H}_B \equiv \mathcal{H}_{B'}$, $\{|i\rangle\}$ denotes the standard basis of $\mathcal{H}_B$.  
\label{Bobs Unlimited Control}
\end{definition}
Like the previous modelling assumption, this is also a choice of physical system (Bob's quantum memory) and of an initial condition for Bob's subsystem (it is maximally entangled with the black hole). As expected for an old black hole, this entanglement is nearly maximal and is here idealised as a collection of EPR pairs. The difference from Page’s approach lies in the operational setting: Bob has collected all previously emitted Hawking quanta in a quantum memory and now has access to their entanglement with the black hole as a resource.

The last three assumptions concerning the initialisation of the qubits together specify the initial state of the entire system, composed of the diary $A$, the black hole $B$, the reference system $R$, and the quantum memory $B'$, to: 
\begin{equation}
   |{\Psi}_{\text{in}}\rangle \equiv |\mathrm{EPR}\rangle_{R A} \otimes |\mathrm{EPR}\rangle_{B B^{\prime}}\,.
   \label{Psiin}
\end{equation}

\subsection{Zooming in: the Procedure behind Information Release}
\label{Details on the Procedure for Information Release}

\begin{figure} 
    \centering
    \includegraphics[scale=0.6]{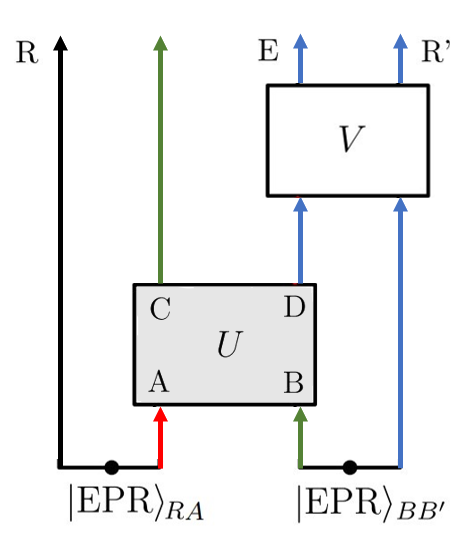}
    \caption{\small Hayden–Preskill QC-model. The quantum diary $A$ (red) is maximally entangled with the reference system $R$ (black). Green lines denote the black hole: $B$
    immediately before the diary is added and $C$ after further evaporation. Blue lines denote systems accessible to Bob: $B'$ represents the relevant subsystem of radiation emitted before diary insertion, and $D$ the newly emitted radiation. $R'$ is a subsystem of $DB'$ of the same size as $A$ where the diary information is reconstructed, and $E$ is its complement. To simplify the figure, lines are weighted according to the number of qubits in the quantum system they represent. The unitary $U$ represents the black hole's scrambling dynamics, and $V$ the decoding operation.
}
    \label{HP}
\end{figure}

The Hayden--Preskill QC-model, illustrated in Figure \ref{HP}, is completely specified through the hypotheses outlined in the previous section. Alice's diary (subsystem $A$), maximally entangled with a reference system (subsystem $R$) of the same size, falls into an old black hole (subsystem $B$). Bob has been collecting
its Hawking radiation in a quantum memory (subsystem $B'$) ever since evaporation started, and, by now, the black hole’s internal state and Bob's quantum memory are in the state $|\mathrm{EPR}\rangle_{BB'}.$

The black hole's unitary dynamics $U_{AB}$ maps the diary and black hole to a state comprising a smaller black hole (subsystem $C$) and new radiation (subsystem $D$). Applied to the initial state $|\Psi_{\text{in}}\rangle$, it produces a pure state $|\Psi_{\text{out}}\rangle$ of the $2(a+b)$-qubit on $RCDB'$:
\begin{equation}
U_{AB}:\mathcal{H}_A\otimes \mathcal{H}_B\to \mathcal{H}_C\otimes \mathcal{H}_D, \quad  
|\Psi_{\text{out}}\rangle
\equiv (I_R\otimes U_{AB}\otimes I_{B'})|\Psi_{\text{in}}\rangle,
    \label{encstate}
\end{equation}
where $|{\Psi}_{\text{in}}\rangle$ was defined in Eq.~\eqref{Psiin}. 
Mathematically, $|{\Psi}_{\text{out}}\rangle$ can be interpreted as a representation of $U_{AB}$ in a doubled Hilbert space via the Choi–Jamiołkowski isomorphism \citep{Jamiołkowski1972linear, CHOI1975completely}. This operator-to-pure state mapping will prove useful for studying the entanglement properties of $U_{AB}$ using standard quantum information measures of entanglement \citep[Sect.~2.1]{hosur2016chaos}.

Bob’s objective is to decode Alice’s quantum state from the information contained in the newly emitted radiation and his quantum memory.\footnote{This does not mean that Bob aims to specify Alice's quantum state in classical bits or directly read its contents, but rather that he wants to transform the quantum state of (part of) the radiation into Alice's state. He can then manipulate the recovered state as he would have been able to if he had direct access to Alice's quantum system.} To reconstruct the diary, Bob has to apply a unitary operation, denoted $V_{DB'}$, acting only on subsystems under his control. Hayden and Preskill do not attempt to construct this decoding operation explicitly. They focus on a simpler question: \textit{whether such a decoding operation exists in principle}.

\subsubsection{Entanglement-Protected Communication}
\label{Entanglement-Protected Communication}

To address this decoding problem, the main insight of Hayden and Preskill is to use the framework of `quantum error correction' in the context of black hole evaporation.  The basic idea of any error correction scheme, quantum or classical, is to \textit{redundantly} encode the information in a larger system so that errors and modifications along the way are not fatal to its successful transmission. The classical strategy of sending a message several times does not work for quantum systems because of the `No-cloning Theorem' \citep{Wootters1982}. {\it Quantum} error-correcting codes rather encode the quantum information into the entanglement structure of a larger system, allowing the original state to be recovered even if some qubits are lost or corrupted.\footnote{Their existence was first demonstrated in the foundational work by \citet{shor1995scheme} and \citet{steane1996error}.}

The errors relevant in the black hole context are `erasures,' defined as the loss of access to a known subset of the physical degrees of freedom. This is because, after encoding the quantum state of the diary across a larger number of degrees of freedom (as in Eq.~\eqref{encstate}), the black hole eventually reveals only a subset of qubits to Bob, namely those in $D$ emitted after further evaporation. The qubits in $C$ retained in the black hole are, in fact, \textit{erased} as far as Bob is concerned. The black hole thus acts as a quantum channel subject to a known subsystem erasure for the transmission of the diary's information.

In their QC-model, the black hole's dynamics is assumed to effectively conceal the information through scrambling, modelled by a random unitary (cf.~\textit{Modelling Assumption \ref{Scrambling Dynamics}}). Is this random unitary a good encoder for the purpose of error correction in a quantum erasure channel?

\begin{figure} 
    \centering
    \includegraphics[scale=0.45]{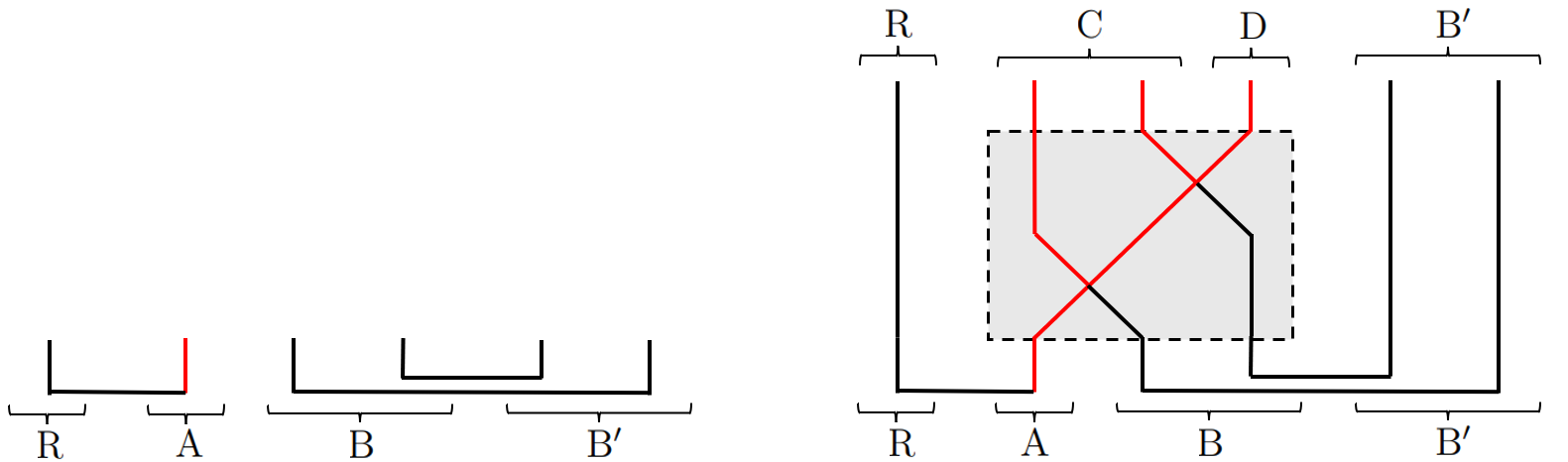}
    \caption{\small Entanglement structure before and after applying the scrambling unitary $U_{A B}$ (dashed box) for a system with a notion of spatial locality. Left: the initial state consists of independent EPR pairs. Right: the resulting multipartite entanglement structure is characteristic of Page scrambling. 
    }
    \label{EntanglementEvolution}
\end{figure}

To answer this question, let us take a look at the effect of applying the scrambling unitary to the input state \citep[Sect.~2.3]{hosur2016chaos}. The initial entanglement structure of the input state $|\Psi_{\text{in}}\rangle$ is one where the composite system $AB'$ is maximally entangled with $RB$, and each individual qubit is maximally entangled with its partner. The information remains localised within the corresponding pairs in $RA$ (see~Figure~\ref{EntanglementEvolution},~left). Once the scrambling unitary $U_{AB}$ is applied to $|\Psi_{\text{in}}\rangle$, this structure transforms into a more complicated arrangement of multipartite entanglement between the reference and output subsystems. The resulting state $|\Psi_{\text{out}}\rangle$ typically has near-maximal entanglement across the relevant bipartitions.
In the terminology introduced above, this is the entanglement structure characteristic of Page scrambling.  This is illustrated in Figure~\ref{EntanglementEvolution} (right).

It is convenient to define the density matrix ${\rho}_{R C D B^{\prime}}$ associated with this output state as
\begin{equation}
    {\rho}_{R C D B^{\prime}} \equiv|{\Psi}_{\text{out}}\rangle\langle{\Psi}_{\text{out}}|\ ,
    \label{encdm}
\end{equation}
and use it to compute the entanglement entropy of arbitrary subsystems. The `mutual information'  is useful in understanding the implications of the entanglement structure of this state for the diary information. For two subsystems $R$, $\widetilde{R}$, it is defined as
\begin{equation}
    I(R:\widetilde{R}) \equiv S_{\text{vN}}(\rho_R)+S_{\text{vN}}(\rho_{\widetilde{R}})-S_{\text{vN}}(\rho_{R\widetilde{R}})\,.
    \label{mutual info}
\end{equation}
This non-negative function quantifies the amount of (classical and quantum) correlations between $R$ and $\widetilde{R}$. In our case, we are interested in the mutual information between the reference system and other parts of the system. If this mutual information is near its maximum, then Alice’s quantum state is approximately recoverable from that subsystem.

In particular, the mutual information between $R$ and a subsystem $X$ of $CDB'$ is approximately zero when their joint subsystem $RX$ is sufficiently smaller than half the size of $RCDB'$. To a good approximation, each entropy term in Eq.~\eqref{mutual info} is individually maximal and $X$ does not reveal any significant information about Alice's diary. However, once $X$ exceeds half the size of $RCDB'$ by a sufficient margin, its mutual information with $R$ approaches the maximum, $2a$, meaning that the diary can be approximately recovered from that subsystem.

In this way, the information about Alice's quantum state, encoded across many degrees of freedom in a highly redundant way, cannot be recovered from sufficiently small subsystems. The corollary is that the erasure of a subsystem, provided it remains approximately uncorrelated with the reference system $R$, does not prevent its approximate recovery from the complementary accessible subsystem. Hayden and Preskill formalise this idea using a simple decoupling argument, described below.

Thus, the black hole's scrambling dynamics is able to protect the information in this quantum erasure channel by entangling it within a larger system.\footnote{
As discussed in Section~\ref{AssumptionsHP}, generating Haar-random unitaries is inefficient, but sufficiently accurate approximate unitary 2-designs provide suitable encoders for protecting the diary against  the relevant erasures and can be generated much more efficiently \citep{dankert2009exact}.} By being ``well hidden'' by the chaotic dynamics, the information in Hayden and Preskill's QC-model is also well protected, a dual aspect of their model the authors describe as a `delicious irony' (\citeauthor{hayden2007blackholes}, \citeyear{hayden2007blackholes}, p.~18).

\subsubsection{Entanglement-Assisted Communication}
\label{Entanglement-Assisted Communication}

Entanglement can be used not only as a means for protecting information, but also as a resource that enables its recovery. The idea is that pre-existing entanglement between the transmitting and receiving ends of the channel can be \textit{used up} to enable more efficient or otherwise impossible forms of communication.\footnote{This idea underlies protocols such as superdense coding \citep{bennett1992communication} and quantum teleportation \citep{bennett1895teleporting}.}

To see how this is possible, we must consider the interplay between entanglement-assisted communication and entanglement-based protection, just discussed. For the scrambling dynamics considered here, Bob can approximately recover the diary only if his accessible subsystem exceeds half of the total system  $RCDB'$. As we discuss now, the assistance provided by pre-existing entanglement allows for this situation to arise relatively quickly.

At the start of the protocol, the old black hole $B$ is maximally entangled with Bob's quantum memory $B'$, as established in \textit{Modelling Assumptions \ref{Old Black Hole}} and \textit{\ref{Bobs Unlimited Control}}. This means that before Alice throws in her diary, Bob already controls half of the system $BB'$. When the diary falls in, one must wait a time of the order of the scrambling time for its information to become delocalised by the black hole's dynamics. The early radiation $B'$ alone nevertheless remains devoid of information about the diary: $I(R:B')=0$. So, how can Bob exploit entanglement to recover the diary?

Hayden and Preskill used a simple decoupling criterion to show that Bob does not need access to the entire system. Their reasoning is as follows: for the diary to be recoverable from $DB'$, the remaining black hole $C$ must have ``lost all memory'' of the diary information content and become decoupled from the reference system, implying that $R$ is now maximally entangled with a subsystem of $D B^{\prime}$. This condition is captured by the approximate factorisation:
\begin{equation}
\begin{aligned}
     {\rho}_{R C} &\approx {\rho}_R \otimes {\rho}_C,
\end{aligned}
    \label{approximatedecoding}
\end{equation}
where closeness is quantified by the decoupling error $\delta$:
\begin{equation}
   || {\rho}_{R C} - {\rho}_R \otimes {\rho}_C||_1 \leq \delta.
\end{equation}
Here, the trace norm $||M||_1=\operatorname{Tr} \sqrt{M^{\dagger} M}$ quantifies the distinguishability of the two states, and the decoupling error bounds the achievable error for the recovery.

For Haar-random dynamics, they estimate that the average trace-norm distance between $\rho_{RC}$ and $\rho_R\otimes\rho_C$ is bounded by a quantity of order $2^{-k}$ once $d=a+k$ qubits have escaped the black hole. Here, $a$ denotes the number of diary qubits, while $k$ is the additional number of emitted qubits that controls the recovery accuracy.\footnote{Sufficiently accurate approximate unitary 2-designs reproduce the relevant estimate, with corrections depending on the design accuracy.} That is, after a relatively small number of extra qubits have been emitted, the diary becomes approximately recoverable by Bob.

In other words, when the size condition
\begin{equation}
   |D||B'|>|C||R|,
   \label{decodingpossible}
\end{equation}
where $|.|$ denotes the Hilbert-space dimension, 
is met for a sufficiently scrambling dynamics, the mutual information between $R$ and $DB'$ becomes nearly maximal. The resulting approximate decoupling guarantees the existence of a decoding unitary $V_{DB'}:\mathcal{H}_D\otimes \mathcal{H}_{B'}\to \mathcal{H}_E\otimes \mathcal{H}_{R'}$ that approximately achieves
\begin{equation}
(I_R\otimes I_C\otimes V_{DB'})|\Psi_{\mathrm{out}}\rangle
\approx |\mathrm{EPR}\rangle_{RR'}\otimes|\phi\rangle_{CE}.
\end{equation}
Here, $R'$ is an $a$-qubit subsystem of the decoder's output (cf.~Fig.~\ref{HP}) that is nearly maximally entangled with $R$, $E$ is its complement, and $|\phi\rangle_{CE}$ is a state of the remaining degrees of freedom.\footnote{The existence of this decoder follows from the decoupling; its efficient implementation is a separate question discussed by \cite{yoshida2017efficient}.} If the diary were instead initially in a pure state $|\psi\rangle$, unentangled with a reference system, the same decoding procedure would approximately recover $|\psi\rangle_{R'}$, reconstructing the quantum information originally stored in Alice's diary.

To summarise this technical analysis of the Hayden–Preskill protocol, our discussion clearly distinguishes between two roles played by entanglement in information release: first, as a means of protecting information against local erasures; and second, as a consumable resource that facilitates the transmission of information.

\subsection{Zooming out: Questions Tackled and their Answers}
\label{Questions Tackled and their Answers}

It is time to zoom out and examine what the Hayden–Preskill protocol teaches us, more generally, about black holes. We recall the new questions their information-theoretic perspective enables us to formulate, introduced in the Introduction:
\begin{itemize}
    \item[(i)] \textit{How does a given amount
of information escape from the black hole?}
    
    \item[(ii)] \textit{How long does Bob have to wait to be able to recover Alice's quantum state with a high degree of confidence?}
\end{itemize}
We now address these questions in turn.

\subsubsection{Physical Mechanism for Information Release}
\label{Physical Mechanism for Information Release}

Within the communication perspective taken by Hayden and Preskill, question (i) can be stated as: `\textit{How does the information get from Alice to Bob in this communication protocol?}' This question has historically led to conceptual puzzles in the context of another communication protocol: quantum teleportation. According to Timpson, (dis)solving these puzzles requires understanding the very nature of `information' \citep[Chap.~4]{timpson2013quantum}. 

Timpson's view of what information is in an information-theoretic setting is based on his understanding of \citet{shannon1948mathematical}'s famous `source coding theorem' and its quantum counterpart, developed by \citet{schumacher1995quantum}. These theorems are usually presented as quantifying how much classical or quantum information a source produces, as measured by the extent to which its output can be compressed. For Timpson, however, they not only provide measures of these quantities but also help identify what is transmitted: `information' is an abstract noun denoting a sequence type of symbols or quantum states produced by an information source \citep[p.~4]{timpson2013quantum}. Such information may be instantiated in various physical systems, but does not itself have a location or path through space and time.

The problems arising from a failure to recognise information as such are evidenced by the answers to the question of how information gets from Alice to Bob during quantum teleportation, which are sometimes surprising. For instance, through the use of backward time travel of classical information \citep{penrose1998quantum, jozsa2003illustrating} or by noting that quantum information can be unexpectedly hidden in classical bits \citep{deutsch2000information}.  If similar answers are given for the Hayden--Preskill protocol, one might wonder whether a proper answer has been given in the first place.

Descriptions of scrambling in the Hayden--Preskill protocol risk a similar confusion. In Section~\ref{Second Model Feature: Circuit Architecture and Size}, we discussed estimates of the black hole scrambling time. Hayden and Preskill did so by envisioning information as a locally conserved fluid and talked about the spreading of the `droplet of information' deposited in the black hole \citep[pp.~11-12]{hayden2007blackholes}. A similar way of talking is employed by Sekino and Susskind, who describe how a system can  return to a scrambled state by suggesting that `the bit of added information will eventually
diffuse over all the degrees of freedom' \cite[p.~4]{sekino2008fast}. Modelling quantum information as a locally conserved fluid or talking about its diffusion imports a spatiotemporal intuition that can be misleading if taken literally. Information, in the sense of \cite{timpson2013quantum} that we adopt, is not a physical substance whose trajectory must be tracked. The physical processes through which information is encoded and transmitted remain, however, subject to causal constraints. Hayden and Preskill themselves recognise that reconciling the rapid spreading of quantum information on the stretched horizon with the black hole’s causal structure might be challenging and, in some sense, deepens the information puzzle \citep[p.~11]{hayden2007blackholes}.

To the extent that this puzzle arises from treating information as a physical substance, it can be addressed similarly to those encountered in quantum teleportation. At their source is the same mistaken assumption: that the word `information' in the question `\textit{How does the information get from Alice to Bob?}'~refers to a concrete spatiotemporal entity whose trajectory we must track. As explained above, in its modern understanding, scrambling is tied to operator growth, through which initially local quantum information becomes delocalised across many degrees of freedom. This should not be understood as the diffusion of a physical substance, but as a change in how the information is physically encoded and from which subsystems it can be reconstructed. Such changes exemplify what Timpson calls `information flow’ \citep[p.~34]{timpson2013quantum}.

Recognising information in this way, Timpson concludes: `The central moral is that one should not be seeking, in an information-theoretic protocol—quantum or otherwise—for some particular `the information’, whose path one is to follow, but rather concentrating on the physical processes by which the information is
transmitted, that is, by which the end result of the protocol is brought about' \citep[p.~5]{timpson2013quantum}. Following this moral, we take the question (i) as an enquiry into the physical processes that enable the transmission of a quantum state from Alice to Bob. Thus, the question is better reformulated as:
\begin{itemize}
  \item[(i')] \textit{What are the physical processes that allow the information to escape?}
\end{itemize}

These processes, as the previous sections demonstrate, include the rapid scrambling of the diary’s quantum information by the black hole’s unitary dynamics, the evaporation of the black hole via Hawking’s process, and, potentially, the collection and decoding of this radiation by a resourceful observer, with entanglement playing the two distinct roles identified above. Together, these processes explain how Alice's information, instantiated in different physical systems, is transmitted to Bob within the Hayden--Preskill model, conditional on its physical and operational assumptions.\footnote{As the protocol involves a sequence of quantum mechanical processes, its physical interpretation is subject to the conceptual challenges of quantum mechanics itself. In particular, how these processes are described will naturally depend on one’s interpretative stance \citep[p.~86]{timpson2013quantum}.}

\subsubsection{Amount of Time before Information Release}

Turning to question (ii), Hayden and Preskill show that the decoding task is information-theoretically possible, i.e.~there exists a unitary decoder $V_{DB^{\prime}}$, provided
\begin{equation}
    d \geq a+k,
    \label{HPresult}
\end{equation}
where $k$ is a constant linked to the decoding error, and where $d$ and $a$ are the number of qubits in the late radiation system $D$ and the diary $A$. Indeed, if there are only $d=a+k$ qubits that are released to Bob, this means that there are $c=b-k$ qubits retained by the black hole, which ensures that the condition in Eq.~\eqref{decodingpossible} is met. What this inequality means is that Bob can recover Alice’s quantum state after waiting for just a few more than $a$ qubits to be emitted.

Once evaporation reaches the half-way point, the delay of information release is thus equal
to the time needed for the black hole to scramble the diary qubits plus the time needed to radiate $a+k$ qubits.\footnote{If Alice had instead thrown her qubits into a ``young” black hole—one not yet highly entangled with its surroundings, so that initially $|B'|\ll|B|$—entanglement-protection would still guarantee that decoding is eventually possible. However, for $DB'$ to satisfy Eq.~\eqref{decodingpossible}, Bob would have to wait for the black hole to emit more than half of its Bekenstein–Hawking entropy, a time of the order of the evaporation time.}
This led Hayden and Preskill to propose the following metaphor: the QC-model of a black hole is hardly black at all; rather, it may be more accurately regarded as an `information mirror' (p.~4), that `returns Alice’s information in a time
comparable to the time it takes the information to fall into the black hole' (p.~12). That is, even though Hawking
radiation escapes slowly and Alice’s diary can potentially contain many degrees of freedom, her secrets, after being
masked for a while, are “reflected” to the exterior.

Yet, we insist that this information mirror has peculiar properties. First, it is a ``deforming mirror'' in the sense that the information decoded or reflected is only an approximation of the original. The recovery error decreases as additional radiation is emitted and, when an approximate unitary design is used, also depends on the accuracy of that approximation. Second, not every observer can see the information being reflected at this fast rate; doing so notably requires access to a resource: pre-existing entanglement shared with the black hole.  Third, seeing this reflection in practice might take much longer, as the time at which the information becomes recoverable in principle is distinct from the time required to implement the decoding operation.

\subsection{Reassessing the Information Controversy}
\label{Information Without Controversy}

Hayden and Preskill's analysis can be summarised as follows: by throwing her quantum diary into an old black hole, Alice is in effect transmitting an arbitrary quantum state through an entanglement-assisted channel involving partial erasure. We now explain that \text{this} use of information avoids the three criticisms concerning its role in fundamental physics raised in Section~\ref{The Information Controversy in Black Hole Physics: an Overview}.

The first critique, formulated by \citet{dougherty2017blackhole}, targeted the epistemic reasoning employed by Bekenstein to explain black hole thermodynamics, specifically the interpretation of the increase in horizon area in terms of an agent’s knowledge of the underlying microphysics. We maintained above, following \citet{wallace2019case}, that Bekenstein’s epistemic motivation has proven unnecessary, since black hole thermodynamics can now be grounded in statistical mechanics through an alternative, physical route (see Sect.~\ref{The Central Dogma of Black Hole Physics}).

What we wish to emphasise, and of which the work of Hayden and Preskill is, in our view, a good example, is that even if information is not essential for grounding black hole thermodynamics, the concept remains significant for other reasons. The diverse degrees of freedom underlying microstate counting in different approaches to quantum gravity can be described abstractly in informational terms. This move proves fruitful by offering a new perspective on the system under study, one that raises novel questions and suggests possible answers under specified physical and operational assumptions.

The second complaint emphasised that the concept of information is inherently linked to its transmission, which presupposes the existence of a communication system. According to \citet{wuthrich2017are}, the description of a black hole provided by general relativity is too impoverished to support the complexity of such a communication system.

In contrast, the theoretical basis on which Hayden and Preskill based their model goes beyond general relativity and incorporates quantum effects. The theoretical results supporting the central dogma are obtained when one considers a theory of gravity coupled to quantum fields. Once the theoretical framework is extended beyond general relativity, black holes reveal themselves to be considerably more complex. 

Under the central dogma and the additional assumptions of the protocol (see Sect.~\ref{AssumptionsHP}), the black hole can be modelled as a quantum communication channel, with the following two characteristics: (1) it is robust against the erasure of any subsystem that is approximately uncorrelated with the reference system, thanks to entanglement-protected encoding; and (2) the successful recovery of the state is enabled by the pre-existing entanglement between Bob's quantum memory and the black hole’s degrees of freedom, allowing for entanglement-assisted decoding.

The third aspect concerns the thesis of informational immaterialism, which claims that all material things are information-theoretic in origin and replaces particles and fields with an immaterial basis of information. Against this thesis, Timpson argues that information is an abstract concept that must be instantiated by something concrete in the material world \citep[Sect.~3.7.1]{timpson2013quantum}. The approach taken by Hayden and Preskill is compatible with Timpson's view: their quantum circuit model provides an abstract informational characterisation of a physical system, namely a black hole as described from the perspective of a distant observer, without treating information as its underlying substance.\footnote{In \cite{de2026holography}, it is argued that, beyond the epistemic and syntactic roles often ascribed to information, notions of information in contexts of quantum gravity of the kind discussed in this paper also have a substantive semantic and physical role.}

\section{Conclusion}
\label{Conclusion}

In this paper, we analysed the Hayden--Preskill model of black hole evaporation, a pioneering example of a
toy model based on quantum circuits and quantum information-theoretic tools that have recently been widely used by theoretical physicists in their investigation of black holes. Our goal was twofold: first, to illustrate that this use of information theory overcomes previously raised scepticism; and second, to assess the strengths and limitations of this approach, which is often relied upon in further investigations.

We have proposed that the particular use of quantum information theory by these authors aligns with the perspective articulated by \citeauthor{timpson2013quantum} (\citeyear{timpson2013quantum}, p.~237) in the conclusion of his book. That is, quantum information theory can serve as a general framework of enquiry, functioning as an alternative to, though not a replacement for, other fundamental theories that describe the behaviour of Nature’s most basic elements, for exploring certain interesting features of material objects. In this sense, Hayden and Preskill investigated the black hole’s properties, such as fast scrambling dynamics and evaporation, for the purpose of communication.

We have argued that the information route to black hole entropy has been rightly criticised by philosophers. Indeed, the thermodynamics of black holes is currently best understood through the physical route, wherein black hole thermodynamics is underpinned by the counting of microstates in partially developed theories of quantum gravity. The present application of quantum information theory does not draw on epistemic reasoning to infer facts about the black hole’s microconstituents. Quite the contrary: such a perspective is only possible once underlying quantum degrees of freedom are assumed.

In the present context, the strength of the quantum information framework is that it is well-suited for formulating and answering quantitative questions about retrieving a given amount of quantum information, i.e.~the quantum state of a number of qubits, from an evaporating black hole. Specifically, Hayden and Preskill focussed on determining when and how such quantum information leaks out in the radiation, and what would be required of an observer to recover it in principle. These are important questions in the context of information loss, which extends beyond the issue of whether black hole evolution is unitary. 

Hayden and Preskill's answer is that, for an old black hole satisfying the model's assumptions, an observer who has access to all the previously emitted Hawking radiation only needs to wait for slightly longer than the time required to emit slightly more qubits than the diary contains. After this delay, the observer can, in principle, recover the original state by performing a (sophisticated) quantum operation on the radiation. This release, relatively rapid in comparison with the evaporation time, occurs because the black hole's dynamics and evaporation allow for quantum error correction combined with entanglement-assisted communication. These are well-established tools in quantum information theory that each make use of entanglement, albeit for different purposes.

The applicability of the Hayden–Preskill model to real black holes depends on the validity of assumptions which, lying at the interface between quantum mechanics and general relativity, are not supported by direct experimental data. At the core of these assumptions lies the central dogma, which provides the necessary rationale for modelling the black hole as a quantum circuit. We discussed how the central dogma has passed non-trivial consistency tests, but it has not been established conclusively to this day. The remaining eight assumptions specify the key features of the circuit model, the communication task, and the resources available for recovering the information. It remains to be seen to what extent these assumptions accurately reflect physical reality, although the assumption that an observer has unlimited control over the radiation is clearly unfeasible for astrophysical black holes.

In any case, as we hope to have illustrated, when addressing precise questions about the retrieval of information, which require a description of black holes beyond that of general relativity, quantum information theory appears to be particularly useful. It provides a precise and usable framework for modelling these systems even in the absence of a fully developed theory of quantum gravity. We expect that the general conclusion, namely that entanglement is crucially involved in the process of information release, will remain valid.

\bibliography{Bibliography}

\end{document}